\documentclass[aps,showpacs,twocolumn, superscriptaddress,prb]{revtex4-1}

\usepackage{graphicx}
\usepackage{amsmath,amssymb,amsfonts}
\usepackage{textcomp}
\usepackage{hyperref}
\usepackage{gensymb}
\usepackage{verbatim}
\usepackage{amsmath}
\hypersetup{breaklinks=true,colorlinks=true,urlcolor=black}
\usepackage{color}
\usepackage{soul}

\begin{document}
	
	\title{Superconductivity versus structural phase transition in the closely related Bi$_{2} $Rh$ _{3.5}$S$ _{2}$ and Bi$_{2} $Rh$ _{3}$S$ _{2}$} 
	
	\author{Udhara S. Kaluarachchi}  
	\affiliation{Ames Laboratory, U.S. DOE, Iowa State University, Ames, Iowa 50011, USA}
	\affiliation{Department of Physics and Astronomy, Iowa State University, Ames, Iowa 50011, USA}
	\author{Weiwei Xie}
	\affiliation{Ames Laboratory, U.S. DOE, Iowa State University, Ames, Iowa 50011, USA}
	\affiliation{Department of Chemistry, Iowa State University, Ames, Iowa 50011, USA}
	\author{Qisheng Lin}  
	\affiliation{Ames Laboratory, U.S. DOE, Iowa State University, Ames, Iowa 50011, USA}
	\author{Valentin Taufour}
	\affiliation{Ames Laboratory, U.S. DOE, Iowa State University, Ames, Iowa 50011, USA}
	\affiliation{Department of Physics and Astronomy, Iowa State University, Ames, Iowa 50011, USA}
	\author{Sergey L. Bud'ko}  
	\affiliation{Ames Laboratory, U.S. DOE, Iowa State University, Ames, Iowa 50011, USA}
	\affiliation{Department of Physics and Astronomy, Iowa State University, Ames, Iowa 50011, USA}
	\author{Gordon J. Miller}
	\affiliation{Ames Laboratory, U.S. DOE, Iowa State University, Ames, Iowa 50011, USA}
	\affiliation{Department of Chemistry, Iowa State University, Ames, Iowa 50011, USA}
	\author{Paul C. Canfield}
	\affiliation{Ames Laboratory, U.S. DOE, Iowa State University, Ames, Iowa 50011, USA}
	\affiliation{Department of Physics and Astronomy, Iowa State University, Ames, Iowa 50011, USA}

	\begin{abstract}
		Single crystals of Bi$_{2} $Rh$ _{3}$S$ _{2}$ and Bi$_{2} $Rh$ _{3.5}$S$ _{2}$ were synthesized by solution growth and the crystal structures, thermodynamic and  transport properties of both compounds were studied. In the case of Bi$_{2} $Rh$ _{3}$S$ _{2}$, a structural first-order transition at around 165\,K is identified by single crystal diffraction experiments, with clear signatures visible in resistivity, magnetization and specific heat data. No superconducting transition for Bi$_{2} $Rh$ _{3}$S$ _{2}$ was observed down to 0.5\,K. In contrast, no structural phase transition at high temperature was observed for  Bi$_{2} $Rh$ _{3.5}$S$ _{2}$, however bulk superconductivity with a critical temperature, $T_{c}\approx$\,1.7\,K was observed. The Sommerfeld coefficient $\gamma$\ and the Debye temperature($\varTheta_{\textrm{D}}$) were found to be 9.41\,mJ\,mol$^{-1}$ K$^{-2}$  and 209\,K respectively for Bi$_{2} $Rh$ _{3}$S$ _{2}$, and 22\,mJ\,mol$^{-1}$ K$^{-2}$  and 196\,K respectively for Bi$_{2} $Rh$ _{3.5}$S$ _{2}$. Study of the specific heat in the superconducting state of Bi$_{2} $Rh$ _{3.5}$S$ _{2}$ suggests that Bi$_{2} $Rh$ _{3.5}$S$ _{2}$ is a weakly coupled, BCS superconductor.
	\end{abstract}
	\maketitle

	\section{Introduction}

		Superconductivity and charge density waves (CDWs) are fascinating and closely linked collective phenomena. The CDW in low dimensional materials was first proposed by Peierls\cite{Peierls1930,Peierls1955}, who showed that a one dimensional metal was unstable against a periodic lattice distortion which creates an energy gap at the Fermi level. Superconductivity and CDW states were  linked when Fr\"{o}hlich proposed a mechanism of superconductivity based on a sliding, incommensurate CDW\cite{Frohlich1954}. Following the formulation of the BCS theory\cite{Bardeen1957} of superconductivity, it was appreciated that the superconducting state and the CDW state are both results of electron-phonon coupling often with the CDW state competing with and ultimately replacing the superconducting state as electron-phonon coupling is increased. In some, relatively rare, compounds both transitions can be found upon cooling; more often, though, a CDW or some other type of structural phase transition removes density of states at the Fermi surface and thus suppresses or even precludes the formation of a superconducting state.

		It is interesting to study the properties of the materials which manifest the  co-existence of superconducting and CDW states, so as to gain a better understanding of how they compete with each other for the density of states as each opens a gap at the Fermi level\cite{Gabovich2000,Gabovich2001,Wilson1974,Harper1975,Monceau1976PRL,DiSalvo1976,Shelton1986PRB,Becker1999PRB,Ramakrishnan2002,Singh2005PRB,Morosan2006Nature}. The electrical transport properties of some ternary, metal-rich chalcogenides\cite{Michener1943,Peacock1950,Brower1974}, $A_{2}M_{3}X_{2}$ ($A$=Sn,Pb,In,Tl and Bi; $M$=Co,Ni,Rh and Pd; $X$=S and Se) have been reported by Natarajan and co-workers\cite{Natarajan1988} with some members of this family showing superconductivity at low temperature\cite{Sakamoto2006,Sakamoto2008,Lin2012}. Recently Sakamoto and co-workers reported that parkerite-type Bi$_{2} $Rh$ _{3}$Se$ _{2}$\cite{Sakamoto2007} was a new superconducting compound (with a critical temperature, $T_c$,\,$\sim$\,0.7\,K) with a possible higher temperature CDW transition at $T_{CDW}$\,$\approx$\,250\,K. Pressure studies on this compound\cite{Chen2014} found that the resistivity anomaly at 250\,K shifted to higher temperature with increasing pressure, which is unusual for a conventional CDW transition\cite{Chu1977PRB,Briggs1980,Budko2006PRB}. Given that isostructural Bi$_{2} $Rh$ _{3}$S$ _{2}$ is reported to have a resistive anomaly near 160\,K (having been measured down to 77\,K)\cite{Natarajan1988}, measurements of single crystalline Bi$_{2} $Rh$ _{3}$S$ _{2}$ to lower temperatures are called for.
		
		Using solution growth out of a Rh rich Rh-Bi-S melt\cite{Lin2012}, our initial growth attempts produced large grain, crystalline material that showed a clear resistivity feature near 160\,K but also an apparent superconducting transition near 2\,K. These results indicated that there may be some form of competition or interaction between structural phase transition and superconductivity in this ternary system. A powder x-ray diffraction measurement revealed the anticipated Bi$_{2} $Rh$ _{3}$S$ _{2}$ phase, but also indicated the presence of a second phase.
		
		In order to better understand the physics and chemistry of this part of the Bi-Rh-S system, in this paper we present details of the crystal growth of both Bi$_{2} $Rh$ _{3}$S$ _{2}$ and a new phase: Bi$_{2} $Rh$ _{3.5}$S$ _{2}$, provide structural data, and present and analyze thermodynamic and transport data from each compound. Whereas Bi$_{2} $Rh$ _{3}$S$ _{2}$ manifests a first order, structural phase transition near 160\,K and does not superconduct for $T>$\,0.5\,K, Bi$_{2} $Rh$ _{3.5}$S$ _{2}$ manifests no signs of any phase transition for 2\,K\,$<T<$\,300\,K has a significantly larger electronic specific heat coefficient, $\gamma$, and manifests bulk superconductivity below 2\,K. Instead of finding the co-existence of a structural phase transition and subsequent superconductivity in one compound we found two closely related compounds: one that manifests a structural phase transition near 160\,K and has a relatively lower electronic specific heat at low temperatures and another that does not undergo a structural phase transition upon cooling, has a relatively larger electronic specific heat and does become superconducting below 2\,K.

	\section{Experimental Methods}
	
		Single crystals of  Bi$_{2} $Rh$ _{3.5}$S$ _{2}$ and Bi$_{2}$Rh$_{3}$S$_{2}$ were produced using solution growth techniques\cite{CANFIELD1992,Canfield2001,Lin2012}. For Bi$_{2} $Rh$_{3}$S$_{2} $, a mixture of elemental Rh, Bi and S was placed in a 2 mL fritted alumina crucible\cite{Petrovic2012} with a molar ratio of Rh:Bi:S\,=\,42.6:28.2:29.2 and sealed in a silica ampule under partial pressure of high purity argon gas. The sealed ampule was heated to 1150\,\celsius\, over  12 hours and held there for 3 hours. After that, it was cooled to 900\celsius\, over 70 hours and excess liquid was decanted using a centrifuge. For Bi$_{2} $Rh$_{3.5}$S$_{2}$, a molar ratio of Rh:Bi:S\,=\,55:22.5:22.5 was used, heated and cooled in a similar manner but slowly cooled to 775\,\celsius\, before decanting. 

		Powder X-ray diffraction data were collected by using a Rigaku Miniflex II diffractometer at room temperature (Cu K$_{\alpha}$ radiation). Samples were prepared by grinding single crystals into powder and spreading them onto a thin grease layer coated single crystal Si, low background puck.  Powder x-ray diffraction data were analyzed using GSAS\cite{Toby2001,GSAS} program.

		Single crystal diffraction data were measured using a Bruker Smart Apex CCD diffractometer\cite{SMART2003} with Mo K$_{\alpha}$ radiation ($\lambda$\,=\, 0.71073\,\AA). Data were collected with mixed $\omega/\phi$ scan modes and with an exposure time of 10\,s per frame. The 2$\theta$ range covered from 6\degree~to 64\,\degree. Intensities were extracted and corrected for Lorentz and polarization effects with the SAINT program. Empirical absorption corrections\cite{Blessing1995} were accomplished with SADABS, which is based on modeling a transmission surface by spherical harmonics employing equivalent reflections with I\,$\textgreater$\,3\,$\sigma$(I). Using the SHELXTL package\cite{Sheldrick2002}, crystal structures were solved using direct methods and refined by full-matrix least-squares on F$^{2}$. Lattice parameters were refined using single crystal diffraction data and are summarized in Table\,\ref{Tb_lattice}. Atomic coordinates and displacement parameters with fully site occupation for Bi$ _{2} $Rh$ _{3}$S$ _{2}$ and Bi$ _{2} $Rh$ _{3.5}$S$ _{2}$ are given in Table\,\ref{Tb_Wyckoff}.

		The ac resistivity ($f$\,=\,17\,Hz) was measured as a function of temperature by the standard four probe method in a Quantum Design, Physical Property Measurement System (PPMS) instrument. Depending on the sample size Pt or Au wires (with the diameter of 25\,$\mu$m or 12.7\,$\mu$m respectively) were attached to the samples using Epotek-H20E silver epoxy or DuPont 4929N silver paint. The speciﬁc heat was measured by using the relaxation method in a Physical Property Measurement System. The $^{3}$He option was used to obtain a measurements down to 0.4 K. The DC magnetization measurements were performed in a Quantum Design, Magnetic Property Measurement System (MPMS).  
		
		\begin{table}[t!]
			\caption{Lattice parameters of Bi$ _{2} $Rh$ _{3}$S$ _{2}$ (293\,K and 120\,K) and Bi$ _{2} $Rh$ _{3.5}$S$ _{2} $ at 293\,K. All values are from single crystal diffraction data. }
			\begin{tabular}{|p{2.5cm}|p{1.8cm}|p{1.8cm}||p{1.8cm}|}
				\hline
				Formula      & Bi$ _{2} $Rh$ _{3}$S$ _{2} $  (293\,K) & Bi$ _{2} $Rh$ _{3}$S$ _{2} $ (120\,K) & Bi$ _{2} $Rh$ _{3.5}$S$ _{2} $ (293\,K)     \\ 
				\hline
				Formula weight & 790.81 & 790.81 & 842.27            \\
				$ Z $-formula units & 4 & 12 & 4 \\
				Space group   & C2/m & C2/m  & C2/m     \\ 
				$a$ ($\textrm{\AA}$) & 11.291(3) & 11.542(2) & 11.5212(3)    \\
				$b$ ($\textrm{\AA}$) & 8.378(2) &  8.341(2)  & 7.9408(2)         \\
				$c$ ($\textrm{\AA}$) & 7.942(4) & 17.768(4)  & 7.8730(3)           \\
				$\beta$ & 133.286(2)\degree & 107.614(2)\degree & 128.033(2)\degree  \\
				$Volume$ ($\textrm{\AA}^{3}$) & 546.8(3)  & 1630.4(5) & 567.33(3)   \\
				$Density$ (g/cm$^{3}$) & 9.605   & 9.664  & 9.861    \\
				
				\hline
				
			\end{tabular}
			\label{Tb_lattice}
		\end{table}
			
	\begin{table}[t!]
		\caption{Atomic coordinates and equivalent isotropic displacement parameters of Bi$ _{2} $Rh$ _{3}$S$ _{2} $ (293\,K and 120\,K) and Bi$ _{2} $Rh$ _{3.5}$S$ _{2} $. All the sites are fully occupied. (U$_{eq }$ is defined as one third of the trace of the orthogonalized U$ _{ij} $ tensor.)}
		\begin{tabular}{|p{.8cm} | p{.8cm}| p{.6cm} | p{1.45cm}| p{1.45cm} | p{1.3cm}| p{1.1cm}|}
			\hline
			\hline
			\centering Atom &	\centering Wyck &	\centering Symm. &	\centering x &	\centering y & \centering	z & $U _{eq} $$\textrm{\AA}^{2}$ \\
			\hline
			\multicolumn{7}{|l|}{Bi$ _{2} $Rh$ _{3}$S$ _{2} $ (293 K)}      \\ \hline
			Bi1 & 4i & m   & 0.0002(1)  & 0      	 & 0.2518(2) & 0.009(1) \\
			Bi2 & 4i & m   & 0.5086(1)  & 0      	 & 0.2596(2) & 0.012(1) \\
			Rh1 & 4f & -1  & 0.25       & 0.25     	 & 0.5       & 0.009(1) \\
			Rh2 & 4i & m   & 0.2472(1)  & 0      	 & 0.2472(2) & 0.015(1) \\
			Rh3 & 4h & 2   & 0          & 0.2411(2)  & 0.5       & 0.013(1) \\
			S   & 8j & 1   & 0.2235(6)  & 0.2704(5)  & 0.187(1)  & 0.010(1) \\ \hline	 
			\multicolumn{7}{|l|}{Bi$ _{2} $Rh$ _{3}$S$ _{2} $ (120 K)}      \\ \hline
			Bi1 & 4i & m   & 0.0709(1)  & 0          & 0.4167(1) & 0.017(1) \\
			Bi2 & 4i & m   & 0.2432(1)  & 0          & 0.2599(1) & 0.018(1) \\
			Bi3 & 4i & m   & 0.4026(1)  & 0          & 0.0830(1) & 0.018(1) \\
			Bi4 & 4i & m   & 0.5778(1)  & 0          & 0.4148(1) & 0.017(1) \\
			Bi5 & 4i & m   & 0.7383(1)  & 0          & 0.2464(1) & 0.018(1) \\
			Bi6 & 4i & m   & 0.9041(1)  & 0          & 0.0831(1) & 0.018(1) \\
			Rh1 & 8j & 1   & 0.3196(2)  & 0.2470(3)  & 0.1674(1) & 0.018(1) \\
			Rh2 & 8j & 1   & 0.4092(2)  & 0.2517(3)  & 0.3347(1) & 0.017(1) \\
			Rh3 & 4i & m   & 0.1550(3)  & 0          & 0.0983(2) & 0.019(1) \\
			Rh4 & 4i & m   & 0.1802(3)  & 0          & 0.5923(2) & 0.019(1) \\
			Rh5 & 4i & m   & 0.4860(3)  & 0          & 0.2499(2) & 0.017(1) \\
			Rh6 & 4h & 2   & 0          & 0.2670(5)  & 0.5       & 0.018(1) \\
			Rh7 & 4e & -1  & 0.25       & 0.25       & 0         & 0.018(1) \\
			S1  & 8j & 1   & 0.0187(6)  & 0.2265(10) & 0.2464(4) & 0.018(1) \\
			S2  & 8j & 1   & 0.1262(6)  & 0.2706(10) & 0.0806(4) & 0.019(1) \\
			S3  & 8j & 1   & 0.3000(6)  & 0.2299(9)  & 0.4232(4) & 0.016(1)	\\ \hline	
			\multicolumn{7}{|l|}{Bi$ _{2} $Rh$ _{3.5}$S$ _{2} $} (293 K)           \\ \hline
			Bi1 & 4i & m   & 0.1896(1)  & 0          & 0.1932(1) & 0.009(1) \\
			Bi2 & 4i & m   & 0.2525(1)  & 0          & 0.7483(1) & 0.008(1) \\
			Rh1 & 4i & m   & 0.0200(1)  & 0          & 0.3371(2) & 0.008(1) \\
			Rh2 & 2d & 2/m & 0          & 0.5        & 0.5       & 0.007(1) \\
			Rh3 & 4g & 2   & 0          & -0.3114(1) & 0         & 0.008(1) \\
			Rh4 & 4e & -1  & 0.25       & 0.25       & 0         & 0.013(1) \\
			S   & 8j & 1   & -0.0031(3) & -0.2889(3) & 0.2908(4) & 0.008(1) \\ \hline
		
			\hline				
		\end{tabular}
		\label{Tb_Wyckoff}
	\end{table}	

	\section{Results}
	
		\subsection{Phases and structures}

\begin{figure}[h!]
			\centering
			\includegraphics[width=1.0\linewidth]{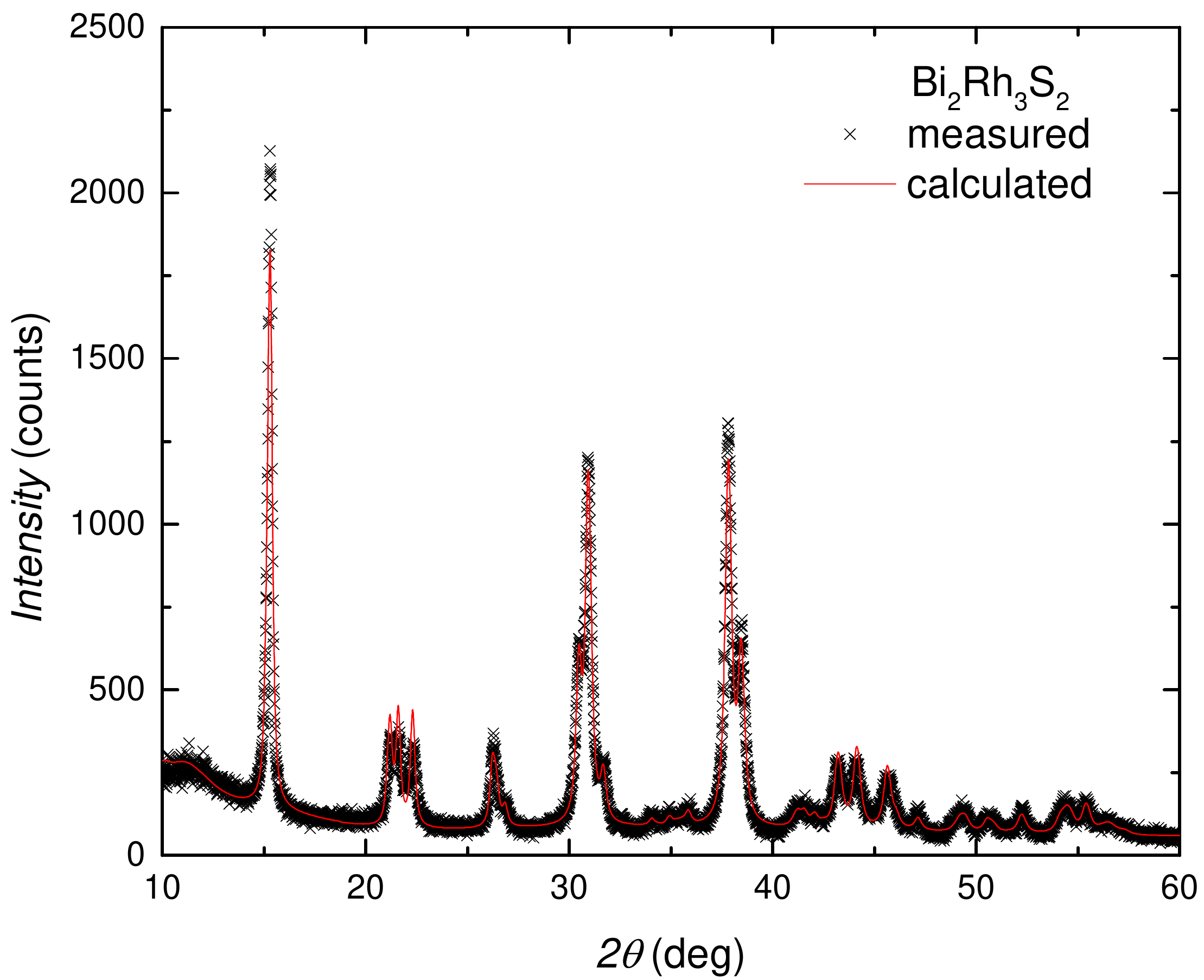}
			\caption{Powder diffraction pattern of pure Bi$ _{2} $Rh$ _{3}$S$ _{2} $. The red line represents the calculated diffraction pattern} 
			\label{Bi2Rh3S2XRD}
		\end{figure}

		The room temperature powder X-ray diffraction pattern from ground, phase pure, single crystals of Bi$ _{2} $Rh$ _{3}$S$ _{2}$ are shown in Fig.\,\ref{Bi2Rh3S2XRD}. Single crystal x-ray diffraction used to refine the lattice parameter of Bi$ _{2} $Rh$ _{3}$S$ _{2}$ with monoclinic C2/m symmetry, $a$\,=\,11.291(3)\,\AA, $b$\,=\,8.378(2)\,\AA, $c$\,=\,7.942(4)\,\AA, and $\beta$\,=\,133.286(2)\degree. These crystallographic parameters are within three standard deviations from literature data of Bi$ _{2} $Rh$ _{3}$S$ _{2}$\cite{Natarajan1988,Anusca2009}. Also these lattice parameters were used to  fit the powder x-ray diffraction data shown in Fig.\,\ref{Bi2Rh3S2XRD}. Resistivity data measured on Bi$ _{2} $Rh$ _{3}$S$ _{2}$ single crystals show only a single transition at\,$\sim$\,165\,K (see Fig.\,\ref{Bi2Rh3S2RT1} below), with no superconductivity being observed down to 0.5\,K.

\begin{figure}[pt!]
			\begin{center}
				\includegraphics[width=1.0\linewidth]{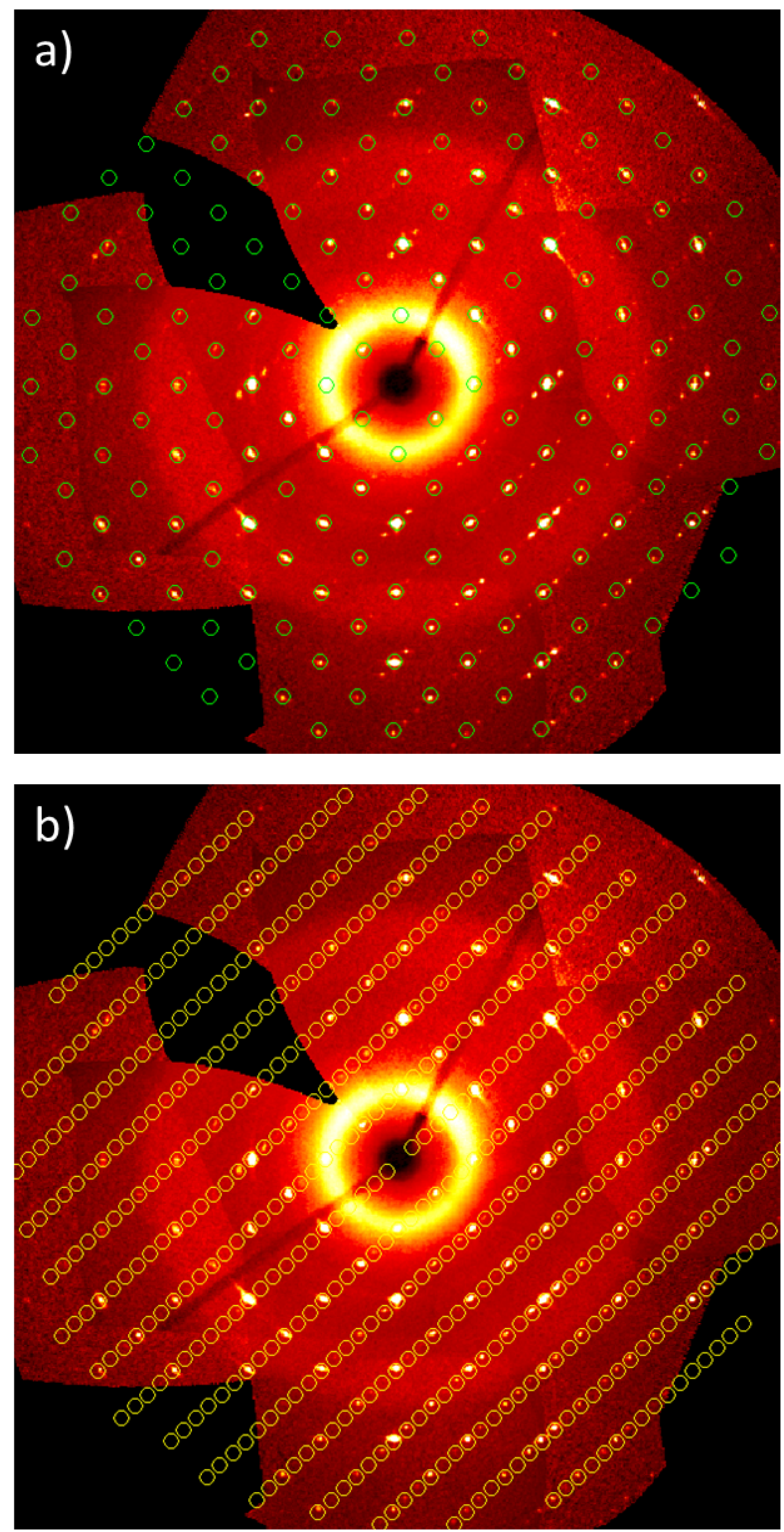}
			\end{center}
			\caption{\label{SingleXtalDiff} Precession images of (h 0 l) zone of a Bi$ _{2} $Rh$ _{3}$S$ _{2}$ single crystal at 120\,K. Green circles in (a) denote reflections that can be indexed by the base cell (C2/m, $a$\,=\,11.291(3)\,\AA, $b$\,=\,8.378(2)\,\AA, $c$\,=\,7.942(2)\,\AA, $\beta$\,=\,133.286(2)\,\degree), note that many reflections can not be indexed. Yellow circles in (b) denote the reflections that can be indexed by the supercell ($a$\,=\,11.542(2)\,\AA, $b$\,=\,8.341(2)\,\AA, $c$\,=\,17.768(4)\,\AA, $\beta$\,=\,107.614(2)\,\degree)} 
			
		\end{figure}

\begin{figure}[th!]
			\begin{center}
				\includegraphics[width=1.0\linewidth]{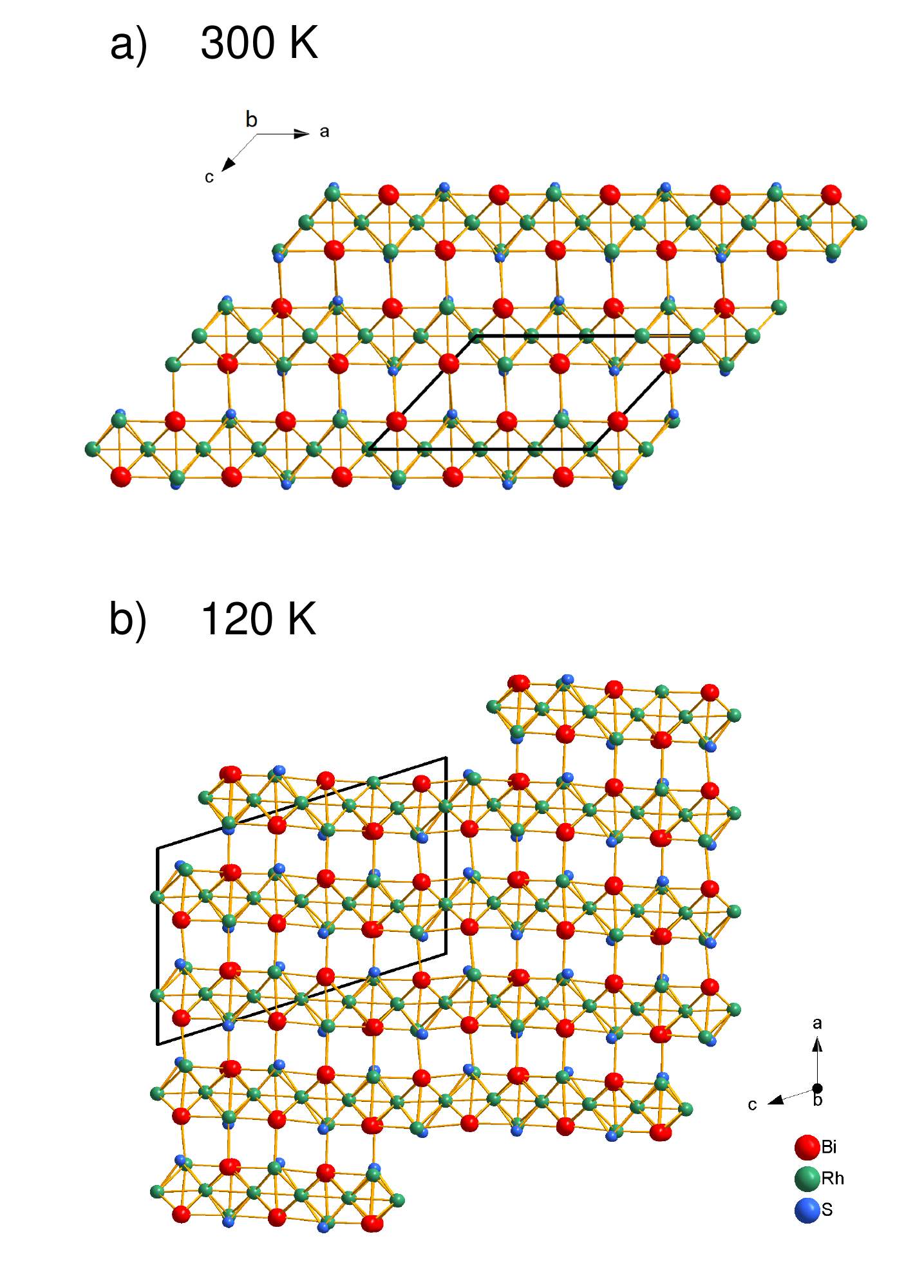}
			\end{center}
			\caption{Crystal structure of Bi$ _{2} $Rh$ _{3}$S$ _{2}$ at 300\,K (a) and 120\,K (b). The black line represents the unit cell. } 
			\label{Bi2Rh3S2_struc}
		\end{figure}

		To understand the phase transition of Bi$ _{2} $Rh$ _{3}$S$ _{2}$ at 165\,K, a set of single crystal X-ray diffraction intensity data of Bi$ _{2} $Rh$ _{3}$S$ _{2}$ was collected at low-temperature (LT), ca.\,120\,K. As shown in Fig.\,\ref{SingleXtalDiff}, extra diffraction spots, not belonging to the unit cell of Bi$ _{2} $Rh$ _{3}$S$ _{2}$ at room-temperature ($a$\,=\,11.291(3)\AA, $b$\,=\,8.378(2)\AA, $c$\,=\,7.943(4)\AA, $\beta$\,=\,133.286(2)\degree), were observed in the (h0l) zone prerecession image. However, all spots could be completely indexed by a larger monoclinic unit cell (C2/m, $a$\,=\,11.542(2)\AA, $b$\,=\,8.341(2)\AA, $c$\,=\,17.768(4)\AA, $\beta$\,=\,107.614(2)\degree), which is about 3 times larger than the unit cell at room-temperature (RT), cf. Table\,\ref{Tb_lattice}. Comparison of the RT and LT structures of Bi$ _{2} $Rh$ _{3}$S$ _{2}$, shown in Fig.\,\ref{Bi2Rh3S2_struc}, indicate that the flat two-dimensional (2D) layers in the RT structure (parallel to ab-plane) are periodically puckered in the LT superstructure. 
		
		\begin{figure}[th!]
			\centering
			\includegraphics[width=1.0\linewidth]{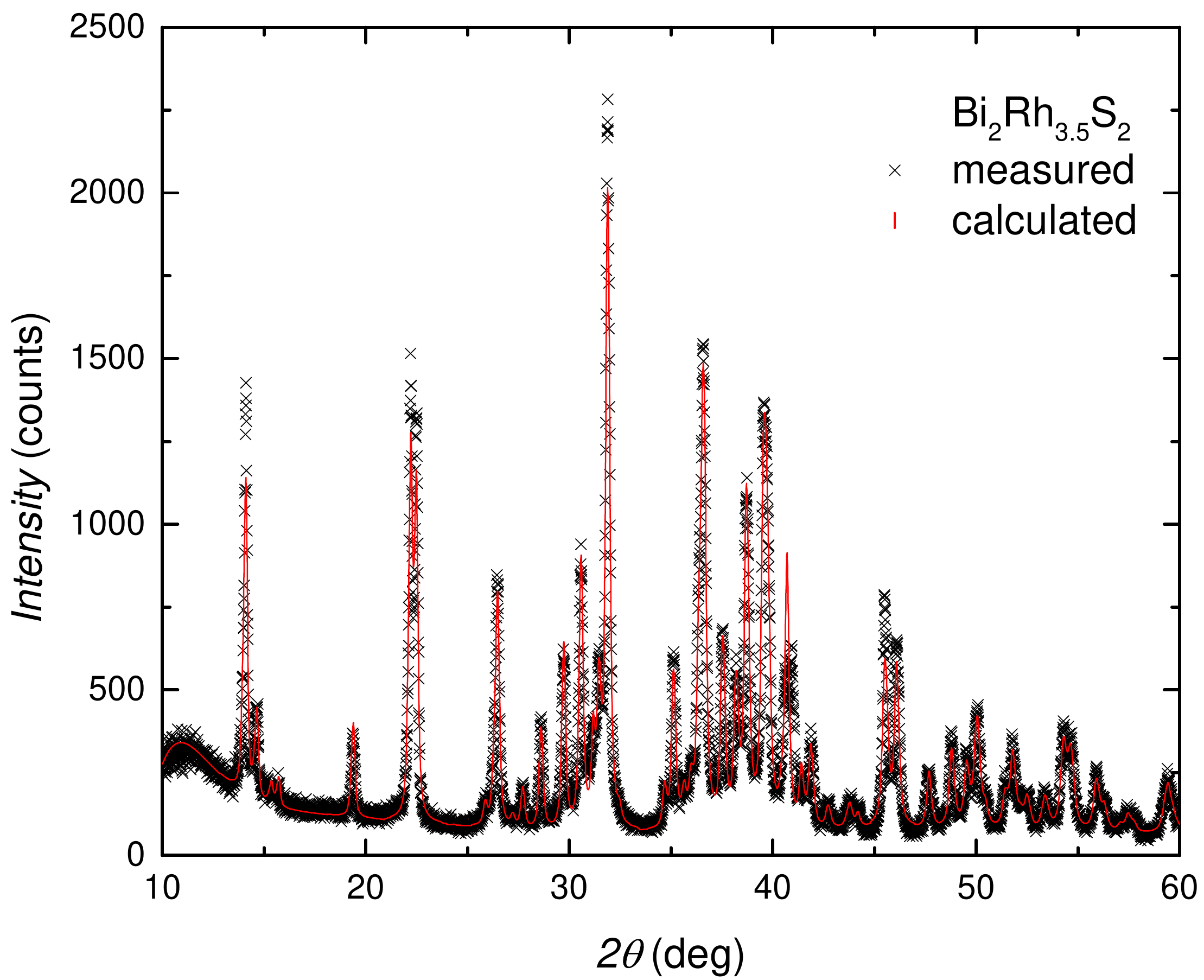}
			\caption{Powder diffraction pattern of Bi$ _{2} $Rh$ _{3.5}$S$ _{2} $. The red line represents the calculated diffraction pattern} 
			\label{Bi2Rh3_5S2XRD}
		\end{figure}
		
		\begin{figure}[th!]
			\begin{center}
				\includegraphics[width=1.0\linewidth]{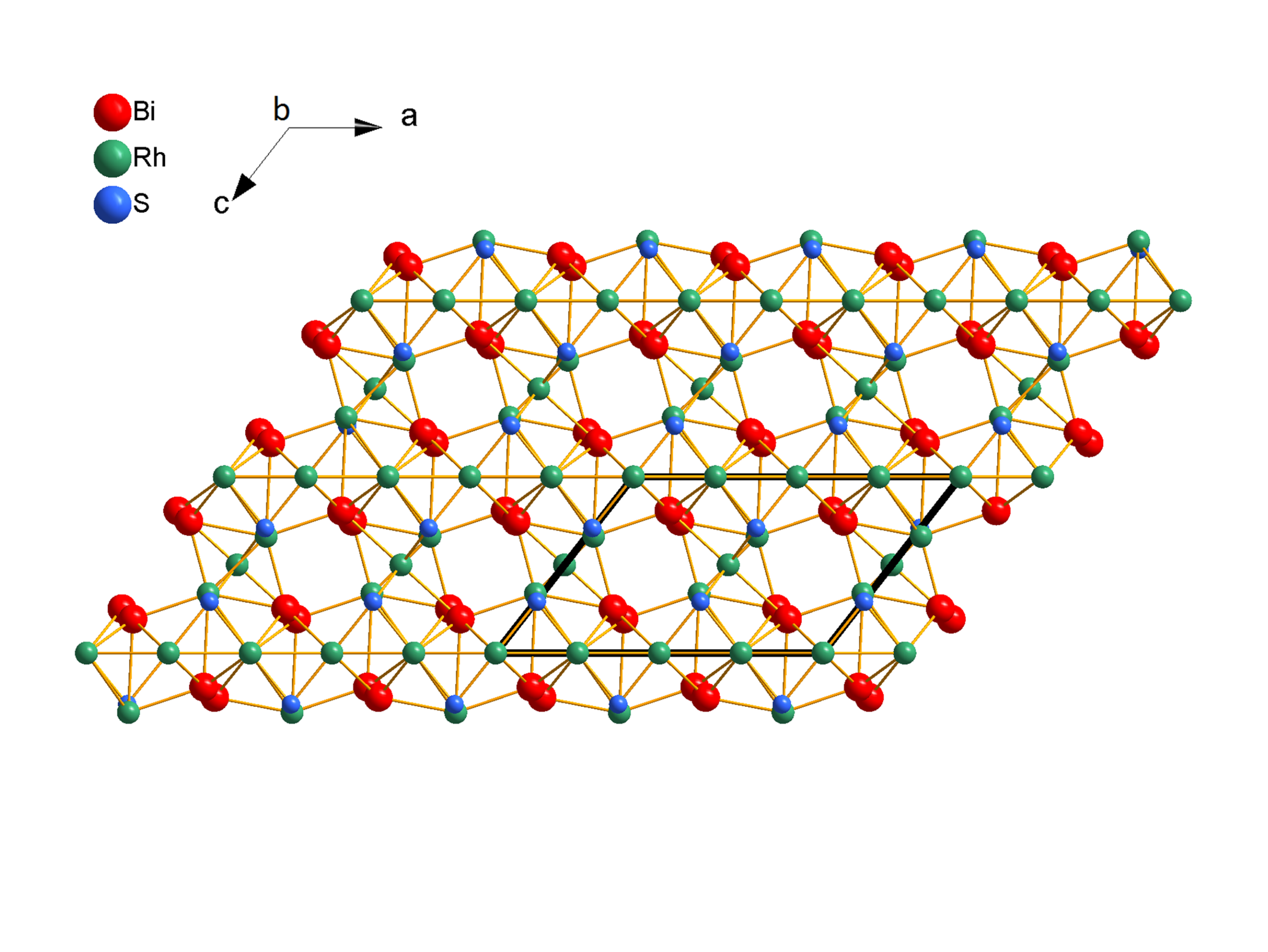}
			\end{center}
			\caption{Crystal structure of Bi$ _{2} $Rh$ _{3.5}$S$ _{2}$. The black line represents the unit cell.} 
			\label{Bi2Rh3_5S2_struc}
		\end{figure}

		The x-ray powder diffraction pattern of a ground, phase pure, single crystal of Bi$ _{2} $Rh$ _{3.5}$S$ _{2}$ is shown in Fig.\,\ref{Bi2Rh3_5S2XRD}. According to single crystal X-ray diffraction analyses (Table\,\ref{Tb_lattice}), Bi$ _{2} $Rh$ _{3.5}$S$ _{2}$ also crystallizes in monoclinic symmetry C2/m ($a$\,=\,11.5212(3)\,\AA, $b$\,=\,7.9408(2)\,\AA, $c$\,=\,7.8730(3)\,\AA, and $\beta$\,=\,128.033(2)\,\degree). These lattice parameters were used to fit the powder x-ray diffraction data shown in Fig.\,\ref{Bi2Rh3_5S2XRD}. Fig.\,\ref{Bi2Rh3_5S2_struc} shows the structure of  Bi$ _{2} $Rh$ _{3.5}$S$ _{2}$ viewed along the (010) direction. Compared to the RT structure of  Bi$ _{2} $Rh$ _{3}$S$ _{2}$, Fig.\,\ref{Bi2Rh3S2_struc}(a), the 2D layers in Bi$ _{2} $Rh$ _{3.5}$S$ _{2}$ are extensively puckered or distorted in response to the insertion of additional Rh atoms into octahedral vacancies between adjacent layers. As will be discussed below, the low temperature electronic specific heat of Bi$_{2}$Rh$_{3.5}$S$_{2}$ is almost double that of Bi$_{2}$Rh$_{3}$S$_{2}$, consist with superconductivity in the former but not the latter.

		\subsection{Physical properties of Bi$ _{2} $Rh$ _{3}$S$ _{2} $}

\begin{figure}[h!]
		\centering
		\includegraphics[width=1.0\linewidth]{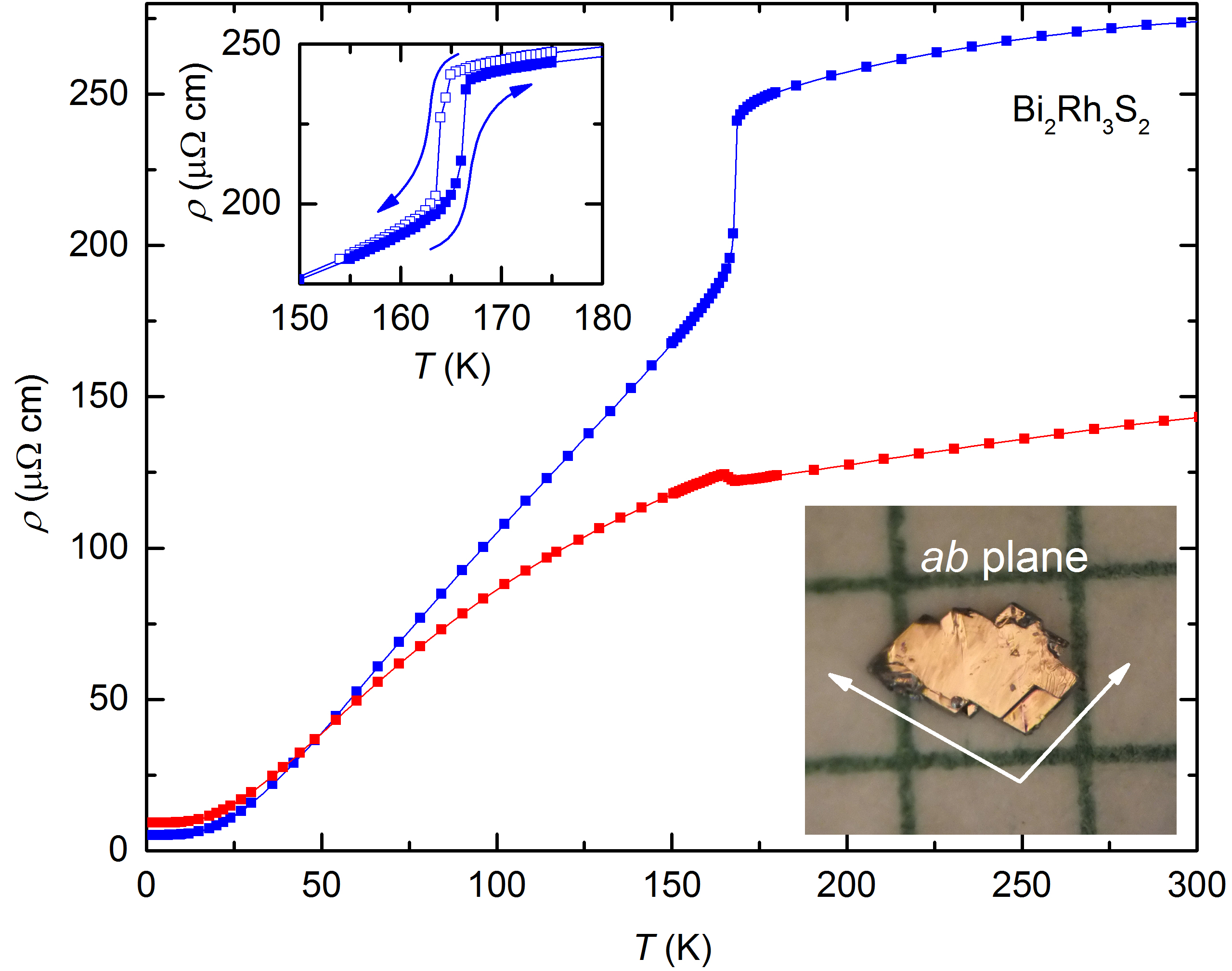}
		\caption{Temperature dependence of resistivity of the phase pure Bi$ _{2} $Rh$ _{3}$S$ _{2} $. The upper inset shows thermal hysteresis observed at 165\,K in resistivity measurement. The lower inset shows a picture of the single crystal of Bi$ _{2} $Rh$ _{3}$S$ _{2} $ on a millimeter grid. Resistivity data were measured on two different samples with current along the directions of white color arrows} 
		\label{Bi2Rh3S2RT1}
	\end{figure}

	The temperature dependent electrical resistivity data from Bi$ _{2} $Rh$ _{3}$S$ _{2}$(Fig.\ref{Bi2Rh3S2RT1}) show a sharp feature associated with a 165\,K transition with no superconducting transition observed down to 0.5\,K. This material shows clear in-plane anisotropy in resistivity. Fig.\,\ref{Bi2Rh3S2RT1} presents the data from two samples with current flowing along each of the edge directions indicated in the lower inset. The residual resistivity ratio (RRR) values are found to be 53 and 15 in these two directions suggesting that the in-plane scattering and/or the Fermi velocity is anisotropic. Above 170\,K the resistivity increases monotonically with temperature, showing metallic behavior. Around 165\,K a sudden increase or decrease of resistivity with decreasing temperature is observed in the two different, in-plane, current directions. Below 160\,K, resistivity again shows metallic like behavior down to 0.5\,K. The upper inset to Fig.\ref{Bi2Rh3S2RT1} shows a 2-5\,K thermal hysteresis observed in the resistivity jump near 165\,K, suggesting a first-order phase transition, as confirmed by our single crystal X-ray diffraction analyses. Whereas this behavior is typical for structural transition, it is less common for a CDW transition\cite{Wilson1974,Monceau1976PRL,DiSalvo1976,Shelton1986PRB,Sakamoto2007,Singh2005PRB,Ramakrishnan2002,Becker1999PRB} which is often second order and usually manifests an increase in resistivity due to a reduction of the density of states at the Fermi energy due to opening up of a gap in the electronic density of states at the Fermi surface. We did not observe $T^{2}$ behavior of the resistivity down to our base temperature.

\begin{figure}[t!]
		\centering
		\includegraphics[width=1.0\linewidth]{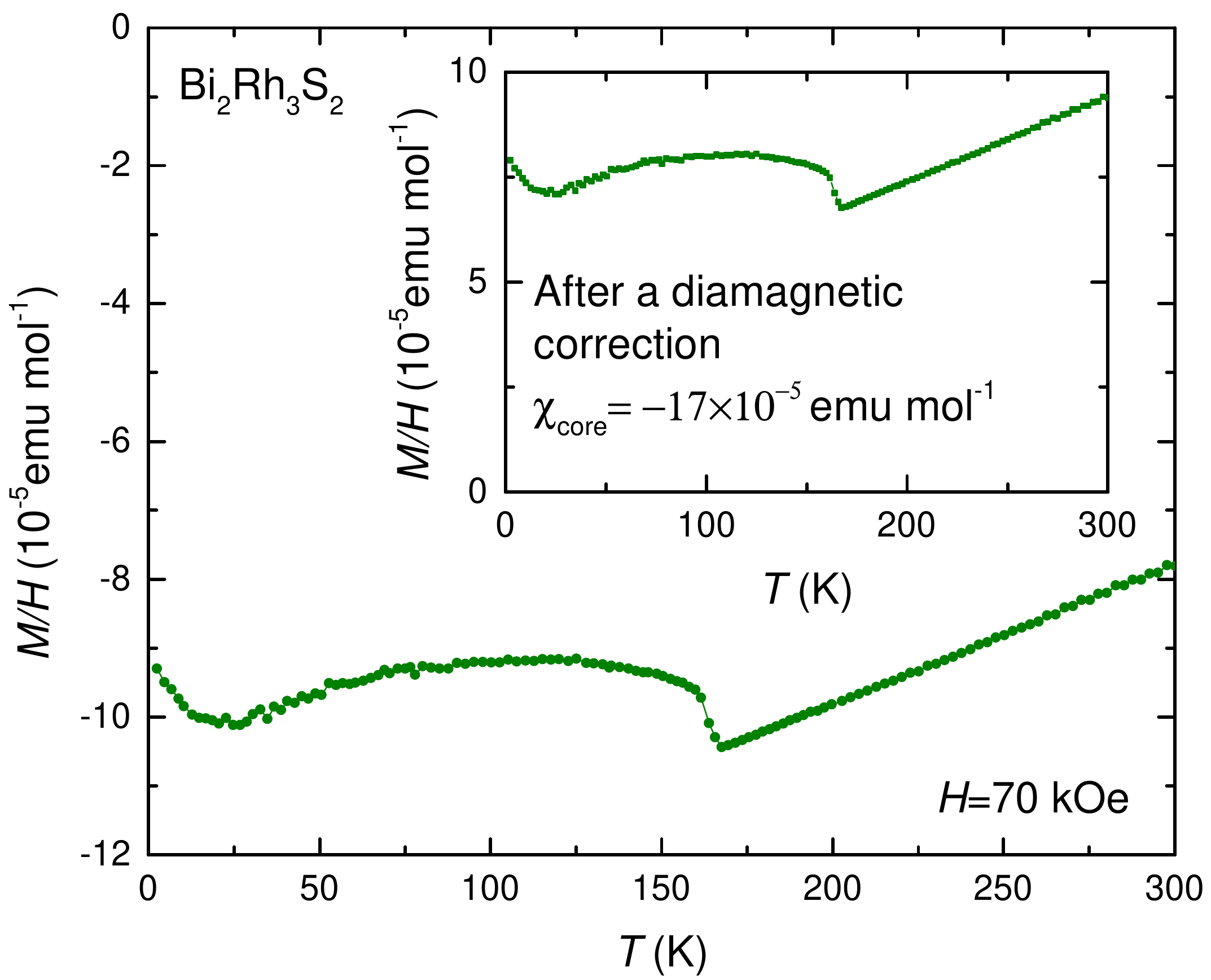}
		\caption{Temperature dependence of susceptibility of phase pure Bi$ _{2} $Rh$ _{3}$S$ _{2} $ before diamagnetic correction ($\chi_{core}$\,=\,-\,17$\times$10$^{-5}$ emu\,mol$^{-1}$)\cite{Mulay1976,Bain2008}. The inset shows after diamagnetic correction.} 
		\label{Bi2Rh3S2MT}
	\end{figure}

		Figure.\,\ref{Bi2Rh3S2MT} shows the temperature dependence of the magnetic susceptibility $\chi\textrm{(T)}$. Since the signal from one piece of single crystal is very low and in order to  measure the magnetic susceptibility, we measured the magnetization of several single crystals encapsulated in a sample holder and subtract the background signal due to the sample holder. The negative value of the magnetic susceptibility indicates the overall diamagnetic behavior of this compound. This is due to the dominating contribution from core diamagnetism. By subtracting the core diamagnetic contribution\cite{Mulay1976,Bain2008} we can estimate the electronic contribution to the susceptibility. The inset of Fig.\ref{Bi2Rh3S2MT} shows the temperature dependence of the electronic contribution to the magnetic susceptibility, after the diamagnetic correction\cite{Mulay1976,Bain2008} ($\chi_{core}$\,=\,-\,17$\times$10$^{-5}$~emu\,mol$^{-1}$). Susceptibility linearly decreases with the decreasing temperature down to 165\,K and then shows dramatic change at 165\,K. Given that both Pauli paramagnetic and Landau diamagnetic susceptibility are proportional to the density of state at the Fermi level $D(\epsilon_{\textrm{F}})$, the change in $\chi$ at 165\,K is consistent with an increase in the density of state at the Fermi level $D(\epsilon_{\textrm{F}})$. In a typical CDW material there is a reduction of the density of state at the Fermi energy due to opening up of a gap in the electronic density of states at the Fermi surface. An increase in density of state is not consistent with the CDW mechanism for structural phase transition. The low temperature upturn (below 25\,K) in the susceptibility is probably due to the presence of a small amount of paramagnetic impurities in the sample.

\begin{figure}[h!]
		\centering
		\includegraphics[width=1.0\linewidth]{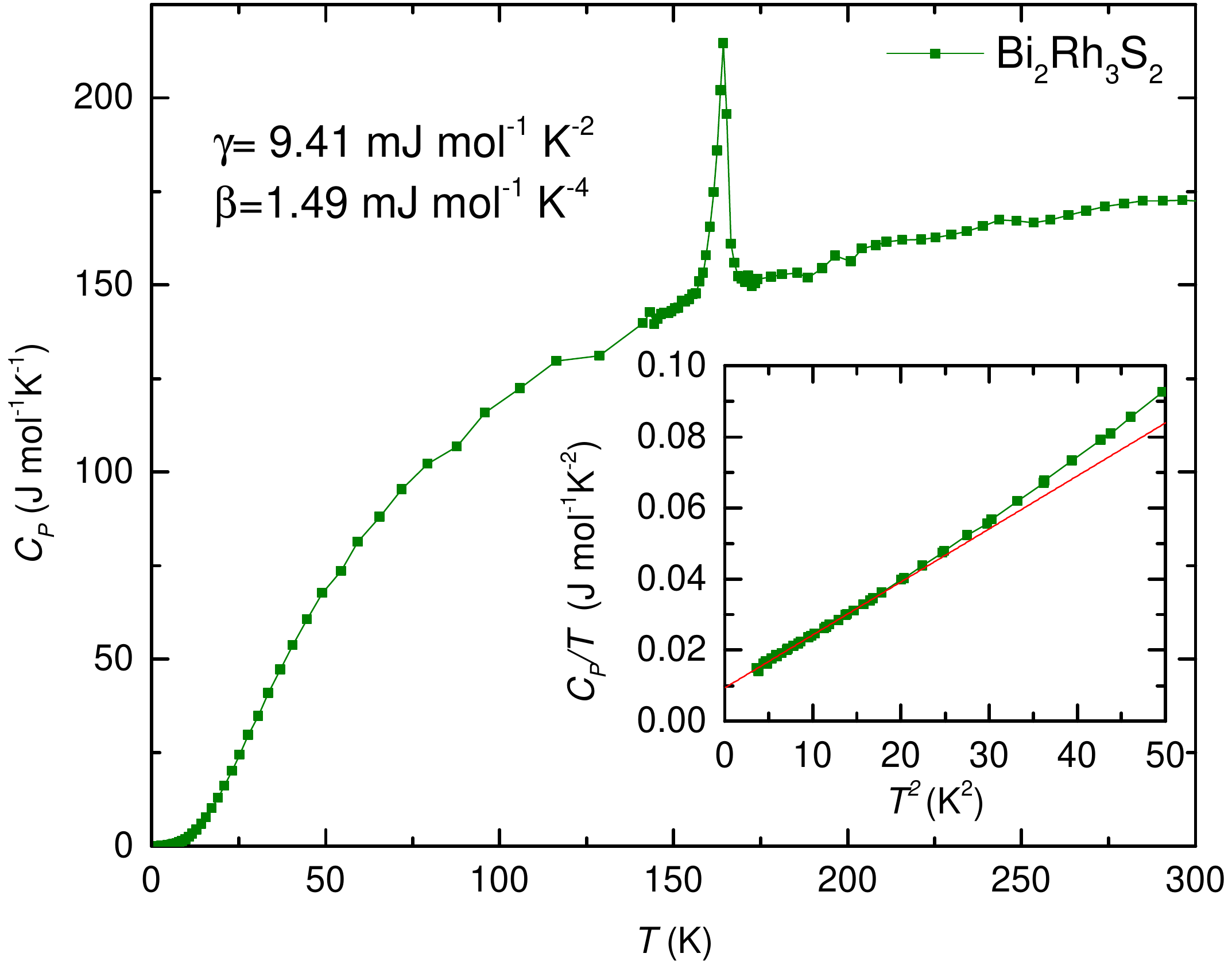}
		\caption{Temperature dependence of specific heat of phase pure Bi$ _{2}$Rh$ _{3}$S$ _{2} $.The lower inset represents the $C_{p}/T$ vs $T^{2}$ graph which used to obtain $\gamma$ and $\beta$ values } 
		\label{Bi2Rh3S2CpT}
	\end{figure}

	The temperature dependent specific heat of Bi$ _{2} $Rh$ _{3}$S$ _{2} $ (Fig.\,\ref{Bi2Rh3S2CpT}) shows a broadened latent heat feature, consistent with the first-order like features seen in the resistivity and magnetization data shown in the Figs.\,\ref{Bi2Rh3S2RT1} and \ref{Bi2Rh3S2MT}. The room temperature specific heat of Bi$ _{2} $Rh$ _{3}$S$ _{2} $, 172.6~J\,mol$^{-1}$\,K$^{-1}$, is close to the Dulong–Petit limit, $C_{V}$=3$nR$=174.6~J\,mol$^-1$\,K$^{-1}$. The Sommerfeld coefficient, $\gamma$\,=\,9.41~mJ\,mol$^{-1}$\,K$^{-2}$ and $\beta$\,=\,1.49~mJ\,mol$^{-1}$\,K$^{-4}$ values were obtained for the LT phase from the low temperature data  fitted with $C_{p}/T=\gamma+\beta T^{2}$ as shown in the lower inset of Fig.\,\ref{Bi2Rh3S2CpT}. From $\beta$ we obtained a Debye temperature ($\varTheta_{\textrm{D}}$) of 209\,K by using equation\,\ref{Eq_DebyeT} and slightly larger than the value of Bi$ _{2} $Rh$ _{3}$Se$ _{2} $. This reflects a higher phonon density of states at low energies which are likely due to the presence of the lighter element "Sulfur" compared to "Selenium". A similar value of $\gamma$=9.5~mJ\,mol$^{-1}$\,K$^{-2}$ was obtained in isostructural Bi$ _{2} $Rh$ _{3}$Se$ _{2} $ compound\cite{Sakamoto2007}.

	\begin{equation}
	\varTheta_{\textrm{D}}=\left( \frac{12\pi^{2}n\textrm{R}}{5\beta}\right)^{1/3}         
	\label{Eq_DebyeT}
	\end{equation}

	With our measurements of single crystal diffraction, resistivity, magnetization and the specific heat, we can conclude the phase transition in Bi$_{2}$Rh$_{3}$S$_{2}$ at 165\,K is a first order structural phase transition. We do not have clear evidence of a CDW being associated with this transition. 
		
	In the case of Bi$_{2}$Rh$_{3}$Se$_{2}$, based on the results of the resistivity, magnetization, thermoelectric power, thermal expansion and low temperature x-ray measurements, Sakamoto et al\cite{Sakamoto2007} conclude that the anomaly at $\sim$\,250\,K is a CDW transition. However, Chen et al\cite{Chen2014} report that, based on their experiments of pressure and selected-area electron diffraction study,  phase transition at $\sim$\,250\,K is not a CDW transition, rather structural transition. In the present case of Bi$_{2}$Rh$_{3}$S$_{2}$ further, advanced measurements would be needed to find any support for a possible CDW.

		
		\subsection{Physical properties of Bi$_{2}$Rh$_{3.5}$S$_{2}$}

\begin{figure}[t!]
		\centering
		\includegraphics[width=1.0\linewidth]{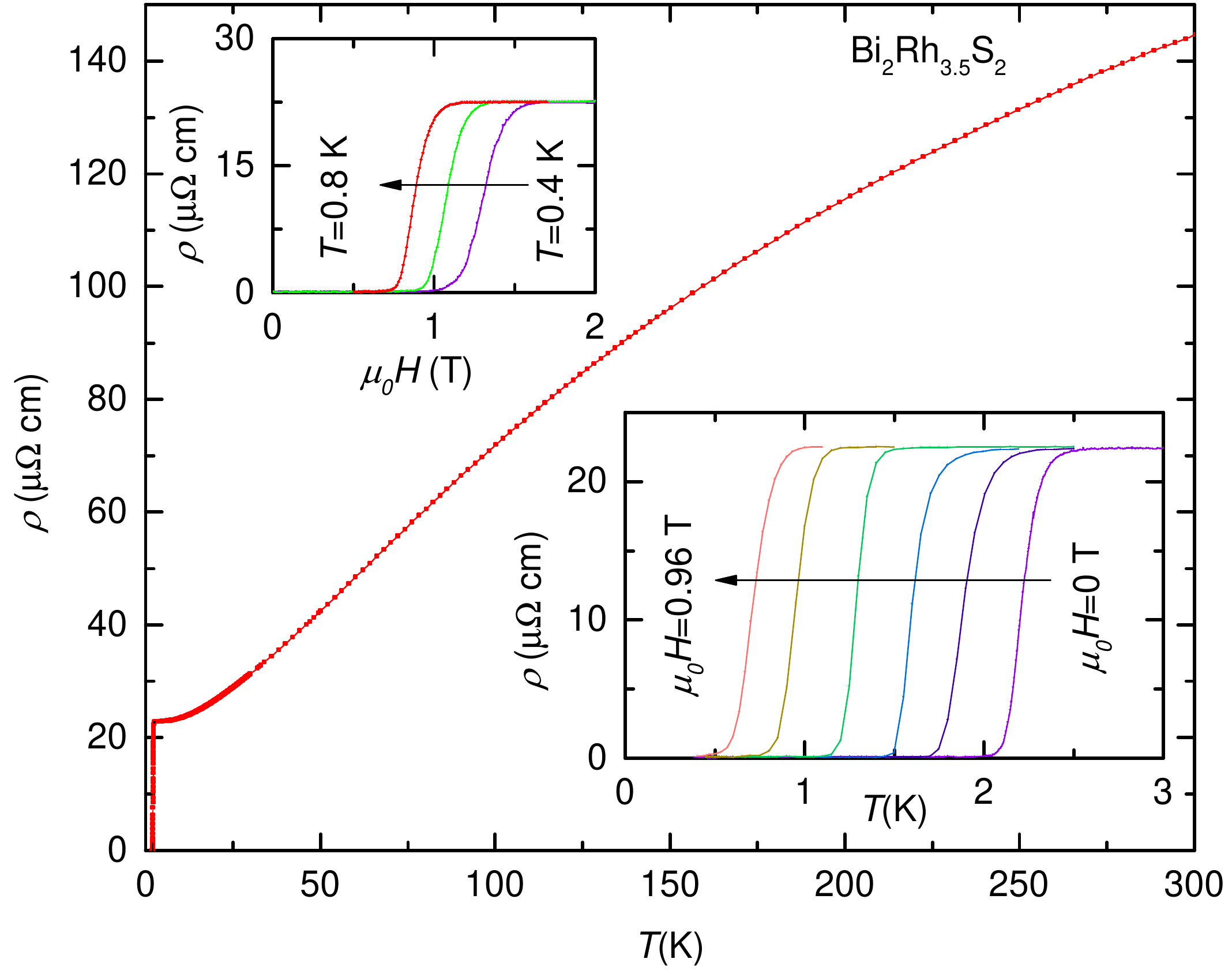}
		\caption{Temperature dependence of resistivity of  Bi$ _{2} $Rh$ _{3.5}$S$ _{2} $. Upper inset shows representative field scans of the resistivity at constant temperatures at 0.4\,K, 0.6\,K and 0.8\,K. Lower inset shows representative temperature scans of the resistivity at constant fields at 0\,T, 0.08\,T, 0.18\,T, 0.42\,T, 0.74\,T and 0.96\,T.} 
		\label{Bi2Rh3_5S2RT}
	\end{figure}

	 Figure.\,\ref{Bi2Rh3_5S2RT} presents the temperature dependence of resistivity of Bi$ _{2} $Rh$ _{3.5}$S$ _{2} $, which, unlike Bi$ _{2} $Rh$ _{3}$S$ _{2} $, does not show any transitions in the resistivity data around 165 K. However, it does manifest zero resistivity below 2\,K, indicating an onset of a superconducting transition at this temperature. RRR of Bi$ _{2} $Rh$ _{3.5}$S$ _{2} $ is 6.4, less than the RRR of Bi$ _{2} $Rh$ _{3}$S$ _{2} $. We did not observe $T^{2}$ behavior in resistivity down to the base temperature of our measurement.

\begin{figure}[t!]
		\centering
		\includegraphics[width=1.0\linewidth]{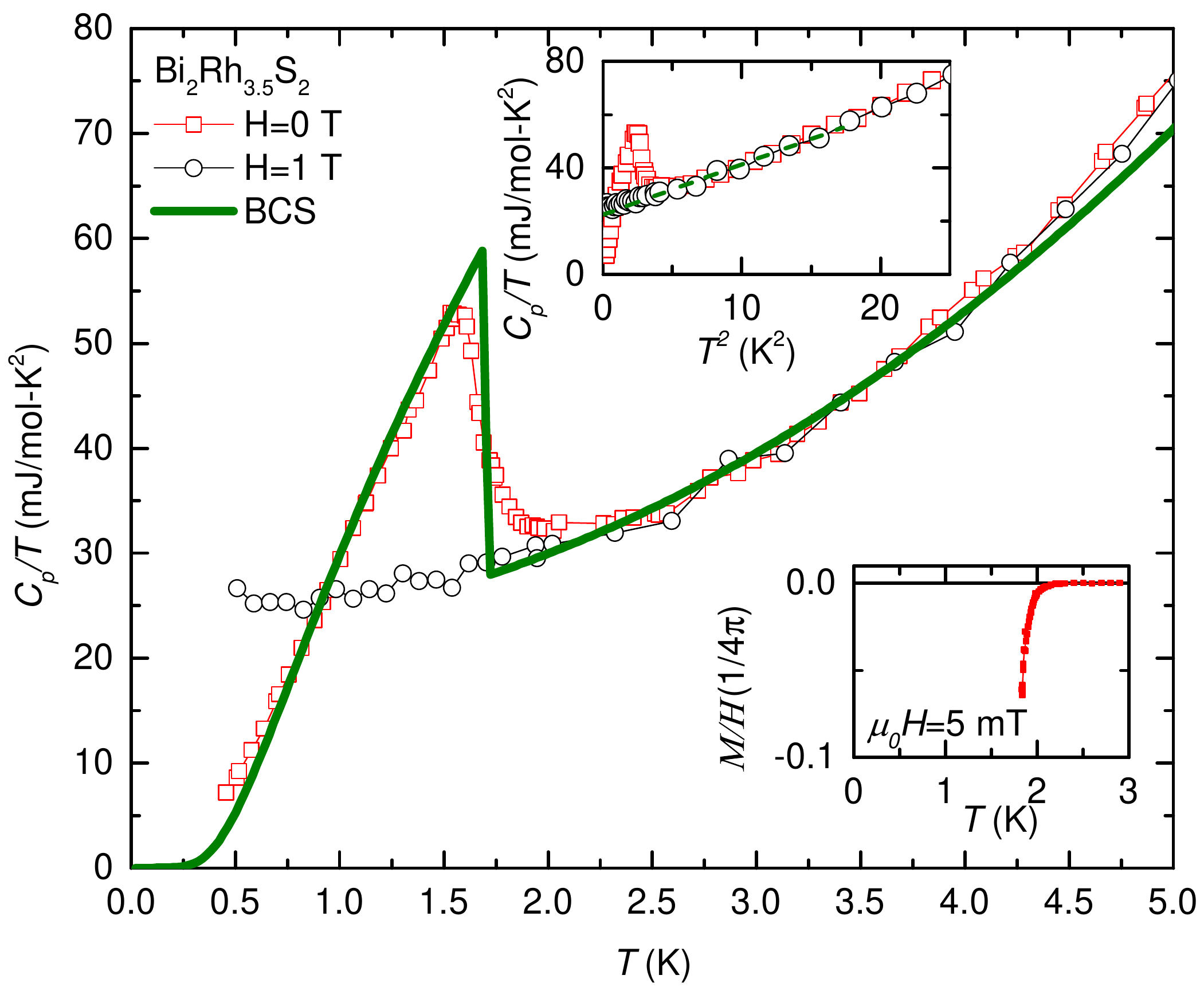}
		\caption{Low temperature $C_{p}/T$ vs T of Bi$ _{2} $Rh$ _{3.5}$S$ _{2} $. Red open squares represent zero field measurements while black open circles represent measurements under 1 T field. Green color solid line shows the BCS calculation. The upper inset represents the $C_{p}/T$ vs $T^{2}$ graph which used to obtain $\gamma$ and $\beta$ values. Lower inset shows ZFC M/H data of Bi$ _{2} $Rh$ _{3.5}$S$ _{2} $.} 
		\label{Bi2Rh3_5S2CP}
	\end{figure}

	Figure.\,\ref{Bi2Rh3_5S2CP} shows the low temperature specific heat data of Bi$ _{2} $Rh$ _{3.5}$S$ _{2} $. The open red squares represent zero field measurements whereas the open black circles represent measurements under 1\,T magnetic field, i.e. $H\geqslant H_{c2}(T)$ for the measured temperature range(see below). From the low temperature data fitted with $C_{p}/T=\gamma+\beta T^{2}$ from 2\,K to 3.8\,K as shown in the upper inset of Fig.\,\ref{Bi2Rh3_5S2CP} we find  $\gamma$\,=\,22\,mJ\,mol$^{-1}$\,K$^{-2}$ and $\beta$\,=\,1.94\,mJ\,mol$^{-1}$\,K$^{-4}$ and can infer that $\varTheta_{\textrm{D}}$\,=\,196\,K. We can clearly see that $\gamma$ of the superconducting Bi$_{2}$Rh$_{3.5}$S$_{2}$ is twice as large as that for the non-superconducting Bi$_{2}$Rh$_{3}$S$_{2}$ or  Bi$_{2}$Rh$_{3}$Se$_{2}$\cite{Sakamoto2007}. Also this value is larger than the other reported parkerite-type superconductors, Bi$_{2}$Ni$_{3}$S$_{2}$\cite{Sakamoto2006}, Bi$_{2}$Ni$_{3}$Se$_{2}$\cite{Sakamoto2006} and Bi$_{2}$Pd$_{3}$Se$_{2}$\cite{Sakamoto2008}.  
	
	On one hand, $T_{c}$\,=\,1.7\,K was obtained by using an equal entropy construction to the low temperature specific heat data. On the other hand, the $H$\,=\,0 $C_{p}(T)$ data start to separate from the $H$\,=\,1\,T $C_{p}(T)$ data below 2.2\,K. Specific heat jump of $\Delta C$\,=\,52.4\,mJ\,mol$^{-1}$\,K$^{-1}$ gives $\Delta C/\gamma T _{c}$\,=\,1.39 which is close to the BCS value 1.43 for a weak-coupling superconductor. The green colored solid line represents a BCS\cite{Bardeen1957,Tinkham1996} calculation. The deviation above 4\,K, normal-state data from the simple Debye model indicates the presence of at least a $T^{5}$ term in the lattice contribution.

	The lower inset of Fig.\,\ref{Bi2Rh3_5S2CP} shows ZFC data of Bi$_{2}$Rh$_{3.5}$S$_{2}$. We were not able to see the full transition in low temperature zero field cooled (ZFC) measurements because the minimum temperature which we can reach in the MPMS is 1.8\,K. These data are consistent with the broadened transition seen in the resistivity and the $C_{p}(T)$ data.

	\begin{figure}[t!]
		\centering
		\includegraphics[width=1.0\linewidth]{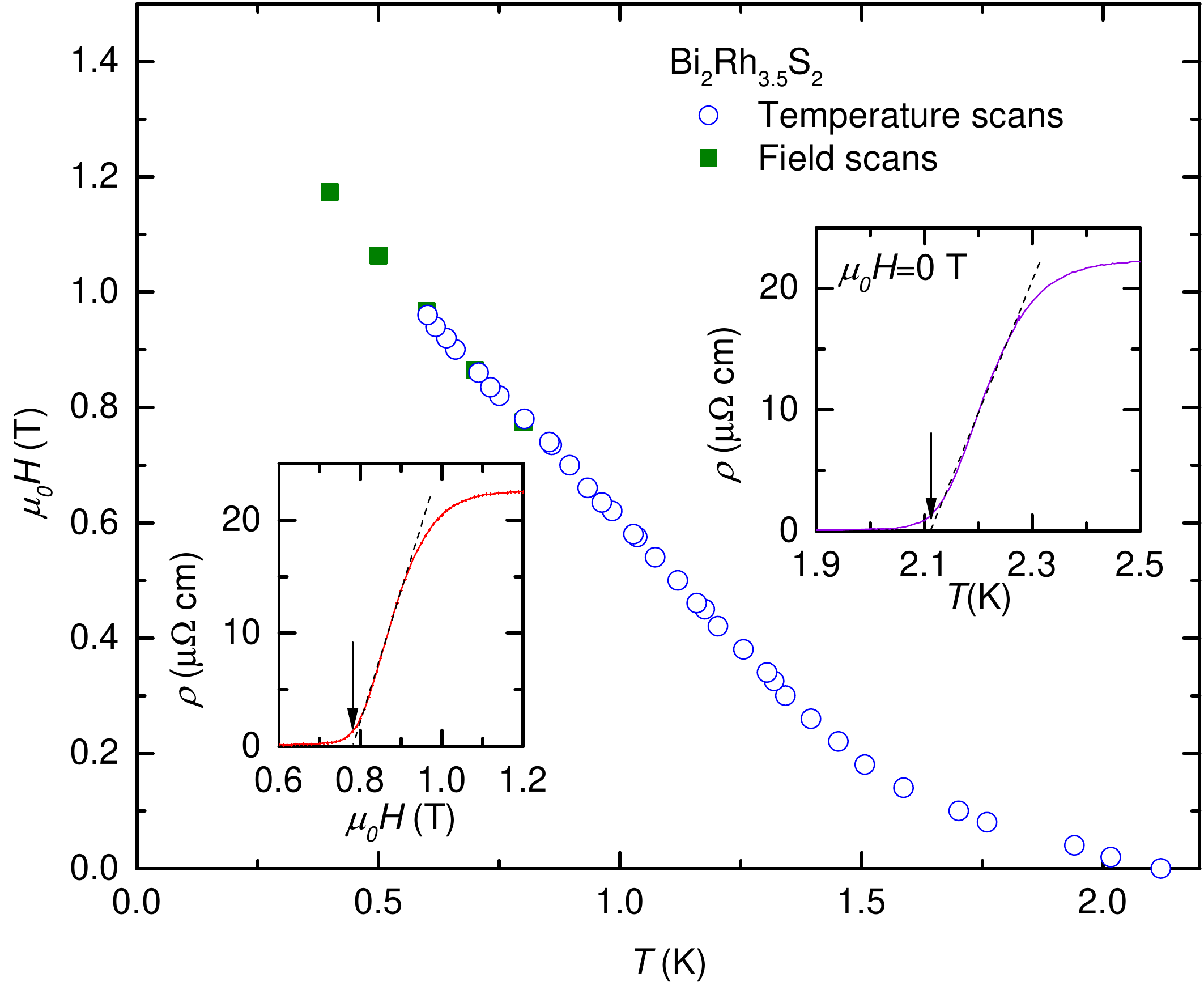}
		\caption{Temperature dependence of the upper critical field for Bi$ _{2} $Rh$ _{3.5}$S$ _{2}$ determined from the resistivity data. Lower and upper insets show the criteria (maximum slope extrapolated to $\rho=0$) which was used to obtain the data points.} 
		\label{Bi2Rh3_5S2HC2}
	\end{figure}
		
	The upper inset of Fig.\,\ref{Bi2Rh3_5S2RT} shows the field dependence of the superconducting transition for temperatures from 0.4\,K to 0.8\,K and the lower inset shows the temperature dependence of the superconducting transition from 0\,T to 0.96\,T applied field. Figure.\,\ref{Bi2Rh3_5S2HC2} shows $\mu_{0}H_{c2}$(T) as a function of the critical temperature determined from the resistivity data. Lower and upper insets show the maximum slope of the resistivity extrapolated to $\rho$\,=\,0, used as a criteria to obtain the $\mu_{0}H_{c2}$ and $T_{c}$ respectively.  We can estimate the Ginzburg-Landau(GL) coherence length\cite{Tinkham1996} at zero temperature, $\xi (0)$=150$\textrm{\AA}$ by using the relation $\mu_{0}H_{c2}$(0)= $\Phi_{0}/{2\pi\xi (0) ^{2}}$, in which $\Phi_{0}$ is the quantum flux and estimating $\mu_{0}H_{c2}$(0) to be 1.5\,T. The value of $\mu_{0}H_{c2}$(0) is well below the  Pauli paramagnetic limit\cite{Clogston1962} of $\mu_{0}H_{\textrm{c2}}^{p}(0)$~=~$1.85\,T_{\textrm{c}}$~=~3.15\,T, suggesting  an orbital pair-breaking mechanism. The upward curvature near $T_{c}$ in the  $H_{c2}(T)$ data shown in Fig.\,\ref{Bi2Rh3_5S2HC2} may come from a distribution of $T_{c}$ values in the sample\cite{Park2008,Xiao2012PRB,Xiao2012PRB86,Slebarski2014PRB} or multi-band superconductivity as was discussed by Shulga et al\cite{Shulga1998PRL} or nonlocal correction\cite{Hohenberg1967PhyRev,Metlushko1997PRL} to the Ginzburg-Landau equations. Similar $H_{c2}$(T) curvatures were found for YNi$_{2}$B$_{2}$C as well as LuNi$_{2}$B$_{2}$C\cite{Shulga1998PRL}.

	The electron-phonon coupling constant $\lambda_{e-ph}$ can be estimated from the McMillan equation\,\cite{McMillan1968} for the superconducting transition temperature (equation\,\ref{Eq_McMillan}), for phonon mediated superconductors,
	
	\begin{equation}
	T_{c}= \frac{\varTheta_{\textrm{D}}}{1.45} \textrm{exp}  \left[  -\frac{1.04(1+\lambda_{\textrm{e-ph}})}{\lambda_{\textrm{e-ph}}-\mu^{*}(1+0.62\lambda_{\textrm{e-ph}}) }\right]       
	\label{Eq_McMillan}
	\end{equation}

	 \noindent where $\mu^{*}$, the Coulomb pseudopotential, having value often between 0.1 and 0.2 and usually taken as 0.13\,\cite{McMillan1968}. Similar values of $\mu^{*}$ have been used in isostructural compounds\cite{Sakamoto2006,Sakamoto2007,Sakamoto2008}. Using $\varTheta_{\textrm{D}}$\,=\,196\,K and $T_{c}$\,=\,1.7\,K we estimated  $\lambda_{\textrm{e-ph}}$\,=\,0.54. A difference of $\mu$ from the assumed  value of 0.13 will give a different value of  $\lambda_{\textrm{e-ph}}$. For example,  $\lambda_{\textrm{e-ph}}$\,=\,0.48 if $\mu$\,=\,0.1 and  $\lambda_{\textrm{e-ph}}$\,=\,0.69 if $\mu$\,=\,0.2. By using both normal state and superconducting state specific heat data, one can obtain the thermodynamic critical field, $\mu_{0}H_{c}$(0) as a function of temperature from equation\,\ref{Eq_ExpHC}

	\begin{equation}
	\frac{\mu_{0} V_{m} H_{c}(\textrm{0})^{2}}{2}=\int\limits_{0}^{T_{c}}\Delta S(T')dT'       
	\label{Eq_ExpHC}
	\end{equation}

	\noindent in which $\Delta S(T)$ is the entropy difference between the normal and superconducting states and $V_{m}$ (8.54$\times$10$^{-5}$ m$^{3}$ mol$^{-1}$) is the molar volume. The calculated value of $\mu_{0}H_{c}$(0) is 23\,mT for Bi$ _{2}$Rh$ _{3.5}$S$_{2}$. This value is larger than the value of Bi$ _{2}$Rh$ _{3}$Se$_{2}$ and clearly reflects the larger $\gamma$ and $T_{c}$ of Bi$ _{2}$Rh$ _{3.5}$S$_{2}$.

 	\begin{subequations}
 	\begin{align}
 	 & \mu_{0}H_{c}(\textrm{0})=1.76\times 10^{-4}\left( 4 \pi  D(\epsilon_\textrm{F})\right)^{1/2} k_{\textrm{B}}T_{c}  \label{Eq_BCSHC_a} \\
 	 & \frac{\gamma}{V_{m}}=\frac{1}{15}\pi^{2}D{(\epsilon_F)}k_{B}^{2}  \label{Eq_BCSHC_b}  \\
 	 & \mu_{0}H_{c}(\textrm{0})=1.76\left(\frac{6\gamma \mu_{0}}{4\pi^{2}V_{m}} \right)^{1/2}T_{c}   
 	\label{Eq_BCSHC_c}
 \end{align}
 \end{subequations}

	\noindent Using BCS theory results\cite{Bardeen1957}(equation\,\ref{Eq_BCSHC_a} and \ref{Eq_BCSHC_b})  we can eliminate the D${(\epsilon_F)}$ term and, for the limit of an isotropic gap function, use equation\,\ref{Eq_BCSHC_c} to calculate the value of $\mu_{0}H_{c}$(0)\,=\,21\,mT. This value is close to the calculated value using specific heat. 
	
	Also the penetration depth $\lambda$(0) and GL parameter $\kappa$ are found to be 7450 $\textrm{\AA}$ and 50  from equations\,\ref{Eq_PenDepth} and \,\ref{Eq_kappa}, respectively.

	\begin{equation}
	\mu_{0}H_{c}(0)\approx\frac{\varPhi_{0}}{2\sqrt{2}\lambda (0) \xi (0)} 
	\label{Eq_PenDepth}
	\end{equation}	

	\begin{equation}
	\kappa\approx\frac{\lambda (0)} {\xi (0)} 
	\label{Eq_kappa}
	\end{equation}		

	\noindent From the specific heat jump and using Rutger's relation, $\Delta C/T_{c}=(1/8\pi\kappa^{2})(\textrm{d}H_{c2}/\textrm{d}T)^{2}|_{T_{c}}$\cite{Welp1989PRL} , one can obtain a similar $\kappa$ value of 30. The obtained  $\lambda$(0) and $\kappa$ values of Bi$ _{2} $Rh$ _{3.5}$S$ _{2}$ are smaller than the values of Bi$ _{2} $Rh$ _{3}$Se$ _{2}$\cite{Sakamoto2007} but the value of $\kappa$ ($\kappa$\,$>$\,1/$\sqrt2$) is large enough  to consider Bi$_{2} $Rh$ _{3.5}$S$ _{2}$ is a type II superconductor.

		\section{Conclusions}

		Single crystals of the closely related Bi$ _{2} $Rh$ _{3}$S$ _{2} $ and Bi$ _{2} $Rh$ _{3.5}$S$ _{2} $ systems were synthesized by high temperature solution growth. Resistivity, magnetization, and specific heat measurements were carried out to characterize their normal states and, for Bi$ _{2} $Rh$ _{3.5}$S$ _{2} $, superconducting properties. Bi$ _{2} $Rh$ _{3}$S$ _{2} $ manifests a structural phase transition around 165 K. No superconductivity was observed down to 0.5\,K. Single crystal diffraction measurements at 120\,K and 300\,K confirmed that a 3-fold superstructure develops. We noticed a large, in-plane, resistivity anisotropy in this compound. Thermal hysteresis and the specific heat feature at 165\,K are consistent with a first-order phase transition. The Sommerfeld coefficient $\gamma$\,=\,9.41\,mJ\,mol$^{-1}$ K$^{-2}$ and the Debye temperature\,=\,209\,K were calculated by using low temperature specific heat data. Based on our measurements, we do not have evidence of a CDW type mechanism being responsible for this transition.

		Bi$_{2} $Rh$_{3.5}$S$_{2}$ adopts a monoclinic (C/2m) crystal structure. This material shows a signature of superconductivity in the resistivity and specific heat measurements consistent with bulk superconductivity below the critical temperature  $T_{c}$ of 1.7\,K. The calculated values for the Sommerfeld coefficient and the Debye temperature are 22\,mJ\,mol$^{-1}$\,K$^{-2}$ and 196\,K respectively. Analysis of the jump in the specific heat at $T_{c}$, suggest that Bi$_{2} $Rh$_{3.5}$S$_{2}$ as a weak electron-phonon coupled, BCS superconductor.

		\section*{ACKNOWLEDGMENTS}
		We would like to thank W.~Straszheim, A.~Jesche, X.~Lin, H.~Kim, E.~Mun, M.~Tanatar for experimental assistance and T.~Kong, S.~Ran, H.~Hodovanets and V.\,G.~Kogan for useful discussions. Part of this work (U.S.K, V.T) was carried out at the Iowa State University and supported by AFOSR-MURI Grant No. FA9550-09-1-0603. Part of this work(W.X,Q.L,G.J.M,S.L.B, P.C.C) was performed at the Ames Laboratory, US DOE, under Contract No. DE-AC02-07CH11358.


\pagebreak
\clearpage

\bibliographystyle{apsrev4-1}
%

\end{document}